\documentclass[twocolomn,showpacs,amsmath,amssymb]{revtex4}
\usepackage{graphicx} 
\usepackage{dcolumn}
\usepackage{bm}

\begin{document}
\title{Plasmon spectrum of two-dimensional electron systems with Rashba spin-orbit interaction}
\author{X. F. Wang}
\email{xuefeng@alcor.concordia.ca}
\address{Department of Physics, Concordia University\\
1455 de Maisonneuve  Ouest, Montr\'{e}al, Qu\'{e}bec, Canada, H3G 1M8}

\begin{abstract}
The dielectric function and plasmon modes of
a two-dimensional electron gas (2DEG) are studied in single- and double-quantum-well structures
with Rashba spin-orbit interaction (RSOI) in the framework of the random-phase approximation.
The RSOI splits each parabolic energy subband of a 2DEG into two nonparabolic spin branches and
affects the electronic many-body correlation and dielectric properties of the 2DEG.
The influence of the RSOI
on the 2DEG plasmon spectrum in single quantum wells appear mainly in three ways:
1) an overall frequency lowering due to the energy band deformation;
2) a weak frequency oscillation stemming from the spin-split energy band;
and 3)an enhancement of the Landau damping as a result of the emerging of the inter-branch
single-particle-excitation spectrum.
In double quantum wells, the above effects are enhanced for the optic plasmon mode but diminished for
the acoustic one.
\end{abstract}


\date{\today} 
\maketitle

\section{ INTRODUCTION}
The growing interest in spintronics is leading to an extensive exploration
of electronic properties of a two-dimensional
electron gas (2DEG) in semiconductor heterostructures with spin-orbit interaction (SOI)
thanks to its promising application in manipulating spins. \cite{dat,wan}
As a relativistic effect of dynamic electrons moving in an electric field, the intrinsic
SOI exists in bulk semiconductors with structural inversion asymmetry,
\cite{dres} while controllable SOI can be introduced by asymmetrically confining a 2DEG
in semiconductor heterostructures to create an average electric field across the electron system.
In general, the former results in
the cubic Dresselhaus term of the SOI and
the latter introduces the Rashba term for isotropic 2DEG and the anisotropic linear Dresselhaus term. \cite{byc,ver,ros}
In some narrow-gap semiconductors, such as
InGaAs,
the Rashba spin-orbit interaction (RSOI) dominates and can be well controlled in the laboratory. \cite{koga}
In the past several years many authors addressed the RSOI effects on one-body ballistic transport
but the study of the RSOI effects on many-body properties of 2DEG is relatively limited.
As is well known, dielectric properties and plasmon spectrum can reveal fundamental many-body correlations in a 2DEG. 
In the presence of the RSOI they have been discussed
in the literature. \cite{chen,maga,ullr,mish,xu}
However, some authors concluded that the charge
plasmon spectrum is not affected by the RSOI \cite{ullr}
and others declared the splitting of the plasmon
spectrum in the presence of RSOI. \cite{maga,xu}

From an experimental point of view, the double-quantum-well
(DQW) structure is a good candidate
for studying the electron correlation and collective modes of
a 2DEG through transport measurements such as Coulomb drag.
In a DQW system composed of two spatially separated quantum wells without
tunnelling, the two 2DEG's, which are coupled with
each other via Coulomb interaction, may oscillate collectively out of phase or in
phase. The former case corresponds to the lower frequency acoustic plasmon
and the latter to the optic plasmon. \cite{das,san,fas,kain,hu}
Both plasmon modes play an important role in the many-body
properties and enhance significantly the Coulomb drag in
bilayer systems. \cite{flen,hill,rojo,wang}
In this paper we focus on the effects of the RSOI on the plasmon modes
in single-quantum-well (SQW) and DQW systems.

\section{Dielectric function}

\begin{figure}[tpb]
\includegraphics*[width=90mm]{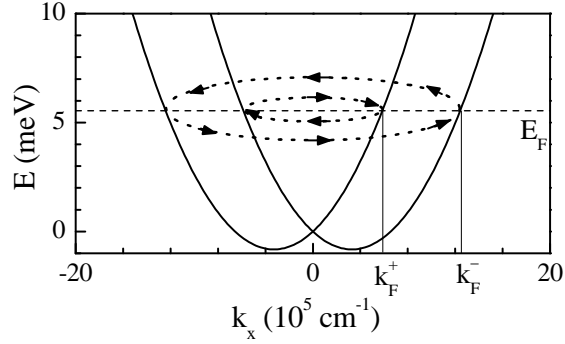}
\vspace{-5cm}
\caption{Energy spectrum of a 2DEG with RSOI. The two dotted ellipses at the Fermi energy $E_F$
illustrate the rotational symmetry in $x$-$y$ plane of the two Fermi circles of the $+$
and $-$ spin branches. The arrows tangent to
the Fermi circles indicate the spin directions.
} 
\label{fig1}
\end{figure}

We consider two Coulomb-coupled 2DEG's without tunnelling in a DQW structure of
a narrow gap semiconductor,
InGaAs/InAlAs,
where the RSOI due to confinement along $z$ direction dominates.
An effective mass $m^\ast=0.05 m_0$, with $m_0$ the free electron mass,
and the background dielectric constant
$\epsilon_\infty=10.8$ are assumed.
In each well the one-electron Hamiltonian of the 2DEG, in the $x-y$ 
plane, is expressed as
\begin{eqnarray}
\hat{H}=\frac{\hat{\bf p}^{2}}{2m^{\ast }}+V(z)+\frac{\alpha}{\hbar}
(\hat{\bf\sigma}\times \hat{\bf p})_{z}.
\end{eqnarray}
The parameter $\alpha$ measures the RSOI strength and is 
proportional to the average interface electric field, $V(z)$ is the electric
confinement potential along the $z$ direction, $\hat{\bf \sigma}=(\sigma 
_{x},\sigma _{y},\sigma _{z})$ denotes the spin Pauli matrices, and 
$\hat{\bf p}$ is the momentum operator.
The eigenstates of the Schr\"odinger equation 
$\hat{H}\Psi({\bf r})=E\Psi ({\bf r})$ are readily obtained as 
\begin{equation}
\Psi ^{\sigma }_{\bf k}({\bf r})=\psi(z)
\frac{e^{i{\bf k}\cdot{\bf r}}}{\sqrt{2}}
\left( 
\begin{array}{c}
1 \\ 
e^{i\phi^{\sigma}_{\bf k}}
\end{array}
\right)
\label{wf}
\end{equation}
with
${\bf k}\equiv(k_x,k_y)$, ${\bf r}\equiv(x,y)$, and $e^{i\phi^{\sigma}_{\bf k}}=\sigma(k_{y}-ik_{x})/k$; $k=\sqrt{k_{y}^{2}+k_{x}^{2}}$ is the radial wavevector
and $\sigma=\pm 1$ denotes the electron spin in the $+$ (upper) or $-$ (lower) branch.
$\psi(z)$ is the wavefunction along the $z$ direction.
$\phi^{\sigma}_{\bf k}$ is the angle between the $x$-axis and the direction of the spin polarization,
which is perpendicular to the electronic wavevector in the
2D plane as shown by the arrows along the Fermi circles in Fig.\ref{fig1}.

The eigenvalues corresponding to the eigenstates $({\bf k},\sigma)$ are
\begin{equation}
E^{\sigma}({\bf k})=\frac{\hbar ^{2}}{2m^{\ast }}k^{2}+\sigma \alpha k.
\end{equation}
This rotationally symmetric electron energy spectrum, $E^\sigma$ vs ${\bf k}$,
is illustrated in Fig.\ref{fig1} for a
strength $\alpha=5\alpha_0$ and an electron density $n_e=1.5n_0$, with $\alpha_0=10^{-11}$ eVm
and $n_0=10^{11}$ cm$^{-2}$. The Fermi wavevector of the spin branch $\sigma$ is 
$k_F^\sigma=\sqrt{2\pi n_e-k_\alpha^2}-\sigma k_\alpha$ with $k_\alpha=\alpha m^\ast/\hbar^2$.

\begin{figure}[tpb]
\includegraphics*[width=90mm
]{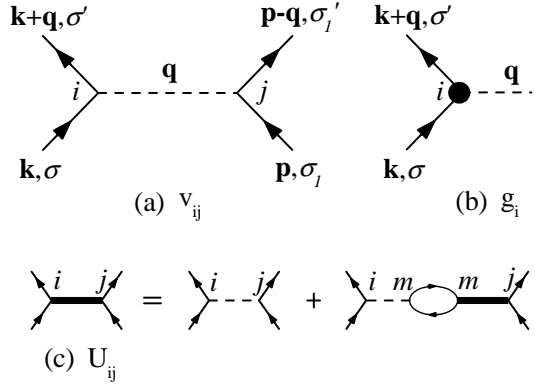}
\vspace{-6cm}
\caption{Diagrammatic representation of (a) the bare Coulomb potential 
$v_{ij,{\bf k},{\bf p}}^{\sigma, \sigma', \sigma_1, \sigma'_1}({\bf q})$,
(b) the spin vertex $g_i^{\sigma, \sigma'}({\bf k},{\bf k}+{\bf q})$,
and (c) the RPA Dyson's equation for the screened Coulomb potential of a 2DEG 
$U^{\sigma, \sigma', \sigma_1, \sigma'_1}_{ij,{\bf k},{\bf p}}({\bf q},\omega)$.} 
\label{fig2}
\end{figure}

Eqs.(1)-(3) apply to each well when the inter-well coupling is neglected and a well index,
$i$ or $j$, will be introduced in the following when the coupling is taken into account.
With the help of Eq.(\ref{wf}), we obtain directly the bare Coulomb interaction between an electron
$({\bf k},\sigma)$
in well $i$ and another one $({\bf p},\sigma_1)$
in well $j$.
As shown diagrammatically in Fig.\ref{fig2}(a), its Fourier transformation reads:
\begin{equation}
v_{ij,{\bf k},{\bf p}}^{\sigma, \sigma', \sigma_1, \sigma'_1}({\bf q})=
g_i^{\sigma, \sigma'}({\bf k},{\bf k}+{\bf q})
v_{ij}(q)g_j^{\sigma_1, \sigma'_1}({\bf p},{\bf p}-{\bf q}),
\label{clbit}
\end{equation}
where $v_{ij}(q)=\int d{\bf r} e^{i{\bf q}{\bf r}}\int dz 
\int dz'|\psi_i(z)|^2 |\psi_j(z')|^2
e^2/(4\pi\epsilon_0\epsilon_i\sqrt{r^2+(z-z')^2})$ is the bare Coulomb interaction between electrons without spin.$\epsilon_0$ is the vacuum dielectric constant and
$\psi_i(z)$ is the wavefunction along the $z$ direction in well $i$, $i=1$ or $2$.
As illustrated in Fig.\ref{fig2}(b), the spin introduces a nontrivial vertex
$g^{\sigma, \sigma'}({\bf k},{\bf k}+{\bf q})=[1+\sigma\sigma' 
e^{i(\phi_{\bf k}-\phi_{{\bf k}-{\bf q}})}]/2$ if the final spin polarization of electron
differs from the initial one
during a Coulomb collision.

The technique for multicomponent system \cite{vin,das} is used to derive
the dynamically screened Coulomb potential in the random phase approximation (RPA).
The following self-consistent equation is found,
as illustrated diagrammatically Fig.\ref{fig2}(c),
\begin{equation}
U^{\sigma, \sigma', \sigma_1, \sigma'_1}_{ij,{\bf k},{\bf p}}({\bf q},\omega)
=v^{\sigma, \sigma', \sigma_1, \sigma'_1}_{ij,{\bf k},{\bf p}}({\bf q})+
\sum_{\substack{\sigma_2, \sigma'_2\\ m,{\bf s}}}
v^{\sigma, \sigma', \sigma_2, \sigma'_2}_{im,{\bf k},{\bf s}}({\bf q})
\Pi^{\sigma_2, \sigma'_2}_{m,\bf s}({\bf q},\omega)
U^{\sigma_2, \sigma'_2, \sigma_1, \sigma'_1}_{mj,{\bf s}-{\bf q},{\bf p}}({\bf q},\omega).
\end{equation}
with
\begin{equation}
\Pi^{\sigma, \sigma'}_{i,{\bf k}}({\bf q},\omega)=
\frac{f_i[E^{\sigma'}({\bf k}+{\bf q})]-f_i[E^{\sigma}({\bf k})]}
{\hbar\omega+E^{\sigma}({\bf k})-E^{\sigma'}({\bf k}+{\bf q})+i\delta}
\end{equation}
and $f_i$ the Fermi distribution function in well $i$.

Note $U^{\sigma, \sigma', \sigma'_1, \sigma'_1}_{ij,{\bf k},{\bf p}}({\bf q},\omega)=
g_i^{\sigma, \sigma'}({\bf k},{\bf k}+{\bf q})U_{ij}(q,\omega)g_j^{\sigma_1, \sigma'_1}({\bf p},{\bf p}-{\bf q})$,
so the above equation reduces to 
\begin{equation}
U_{ij}(q,\omega)=v_{ij}(q)+\sum_{\substack{\sigma_2, \sigma'_2\\m, {\bf s}}}
v_{im}(q)g_m^{\sigma_2, \sigma'_2}({\bf s},{\bf s}-{\bf q})
\Pi^{\sigma_2, \sigma'_2}_{m,\bf s}({\bf q},\omega)
g_{m}^{\sigma'_2, \sigma_2}({\bf s}-{\bf q}, {\bf s})
U_{mj}(q,\omega).
\end{equation}
The elements of the corresponding $2\times 2$ RPA dielectric matrix $\hat{\epsilon}$, defined by 
$v_{ij}(q)=\sum_m \epsilon_{im}(q,\omega)U_{mj}(q,\omega)$, reads
\begin{equation}
\epsilon_{ij}(q,\omega)=\delta_{ij}-v_{ij}(q)\sum_{\substack{\sigma, \sigma'\\j, {\bf k}}}
\frac{1}{2}(1+\sigma\sigma'\frac{k+qcos\theta}{|{\bf k}+{\bf q}|})
\Pi^{\sigma, \sigma'}_{j,\bf k}({\bf q},\omega)
\label{diel}
\end{equation}
with $\theta$ being the angle between ${\bf k}$ and ${\bf q}$.

\section{Plasmon dispersion}
The inverse of the dielectric matrix, $\hat{\epsilon}^{-1}$
describes the many-body response to the two-body
bare Coulomb interaction $v_{ij}$. At the poles of $\hat{\epsilon}^{-1}$,
i.e. for a set value of $({\bf q},\omega)$
a particular element of $\hat{\epsilon}^{-1}$ becomes singular, a density fluctuation of the 2DEG can lead to
a density oscillation in the system or the onset of a collective mode.
It is well known that the zeros of the real part of the
determinant of $\hat{\epsilon}$ describes the plasmon dispersion and the imaginary part the damping of plasmon.
In this paper, we discuss the plasmon spectrum by solving the equation $Re|\epsilon(q,\omega)|=0$ for
the 2DEG in
a SQW and a DQW in the presence of RSOI.

\subsection{Plasmons in a single well}
If the coupling between 2DES's in different quantum wells is negligible,
each well is independent and can be treated as a SQW.
The well indices $i$ and $j$ drop and the dielectric matrix reduces to a scalar function
\begin{equation}
\epsilon(q,\omega)=1-v(q)\hat{\Pi}({\bf q},\omega)
\label{dielsg}
\end{equation}
with the nointeracting correlation function
$\hat{\Pi}({\bf q},\omega)
=\sum_{\sigma, \sigma'}\hat{\Pi}^{\sigma, \sigma'}({\bf q},\omega)
=\sum_{\sigma, \sigma', {\bf k}}
[1+\sigma\sigma'(k+qcos\theta)/|{\bf k}+{\bf q}|]
\Pi^{\sigma, \sigma'}_{\bf k}({\bf q},\omega)/2.$
Although the RSOI splits the energy bands of different spins,
the many-body Coulomb interaction between spin branches is described by
a scalar dielectric function in the RPA rather than a matrix as in confinement induced
multisubband systems. \cite{vin} 
The reason is that the RSOI does not change the wavefunction along $z$
direction and the spin vertex is
separable from the no-spin Coulomb interaction as shown in Eq.(\ref{clbit}).
Eq. (\ref{dielsg}) coincides with the dielectric function obtained in Ref. \onlinecite{chen} and
\onlinecite{mish} but not with Eq. (2) in Ref. \onlinecite{xu}.

In the absence of SOI, the zero-temperature plasmon dispersion relation,
frequency $\omega_0$ vs wavevector $q$,
is approximately expressed as
\begin{equation}
\omega_0^2=\frac{\hbar^2q^2\Delta^2}
{4(m^\ast)^2(\Delta^2-1)}[4k_F^2+q^2(\Delta^2-1)]
\label{plsm}
\end{equation}
with $\Delta=1+\pi\hbar^2/[m^{\ast}v(q)]$ and the Fermi wavevector $k_F=\sqrt{2\pi n_e}$.
In the long wavelength limit $q\ll k_F$, the above equation reduces to the well-known
zero-temperature 2D plasmon frequency
$\omega_p=[n_ee^2q/(2\epsilon_0\epsilon_im^\ast)]^{1/2}$ if the thickness of the quantum well is
negligible.

In the presence of RSOI and $q\ll k_F^\sigma$, the intra-branch correlation function is given
approximately by
\begin{equation}
\hat{\Pi}^{\sigma,\sigma}({\bf q},\omega)=
(1+\sigma k_\alpha/k_F^\sigma)q^2(k_F^\sigma)^2/(2\pi m^\ast \omega^2).
\label{intra}
\end{equation}
The second term is a result of
the deviation of the energy band from parabola. Since $k_F^- > k_F^+$, this term leads to an extra
negative term for the total intra-branch correlation $\hat{\Pi}^{++}+\hat{\Pi}^{--}$ and
lowers the frequency of the plasmon.
The inter-branch correlation is negligible in the long wavelength-limit because electron spins
of the same wavevector in different branches are opposite.
The long-wavelength 2D plasmon frequency with RSOI then also has a very simple dependence
on the system parameters,
\begin{equation}
\omega_p^\alpha = \omega_p[1-k_\alpha^2/(4\pi n_e)].
\label{ratio}
\end{equation}
In the general case the plasmon dispersion relation  is obtained numerically by searching
the zeros of the real part
of Eq.(\ref{dielsg}).
For the sake of simplicity, only electrons in the lowest confined subband are considered and
the electron wavefunction along the $z$ direction is approximated by that
of an infinite highly square-well potential with width
$b=23$\AA, which is valid if the quantum well is not too wide.
The bare Coulomb potential is then
$v(q)=e^2F(qb)/(2\epsilon_0\epsilon_i q)$
with $F(\xi)=\{3\xi+8\pi^2/\xi-32\pi^4(1-e^\xi)/[\xi^2(\xi^2+\pi^2)]\}/(\xi^2+\pi^2)$.

\begin{figure}[tpb]
\vspace*{-4cm}
\includegraphics*[width=100mm]{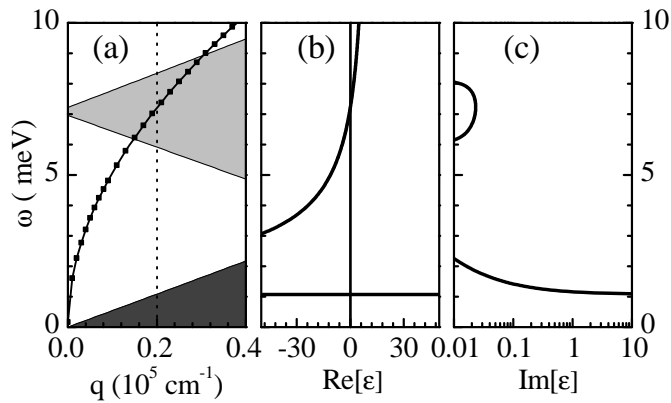}
\vspace{-3cm}
\caption{(a) Plasmon spectrum of a 2DEG of density $n_e=20n_0$ in a SQW with SOI of strength
$\alpha=\alpha_0$ (solid curve) and in a SQW without SOI (filled squares). Inter- (darker shaded) and intra- (light shaded) branch
 electron-hole continuums in the presence of SOI are also shown. The vertical dashed line
indicates the cross-section of the spectra at $q=0.2\times 10^{-5}$ cm$^{-1}$.
(b) The real part and (c) the imaginary part of the dielectric function $\epsilon$ vs energy $\omega$ at $q=0.2\times 10^{-5}$ cm$^{-1}$. Limited by the space,
the segments of the curves for Re$[\epsilon]<-40$ in the energy range $(1,3)$ meV,
Re$[\epsilon]>40$ in $(0,1)$,
$0 < $Im$[\epsilon]<0.01$ in $(2.2,6.2)$ and $(8,10]$, and $10<$Im$[\epsilon]$ in $(0,1.1)$
are not shown. The temperature is $T=2$ K.
} 
\label{fig3}
\end{figure}

In Fig. \ref{fig3}(a) we plot the plasmon spectrum and the single continuum
of a 2DEG with RSOI in an
InGaAs/InAlAs SQW of typical parameters appeared in recent experiments. \cite{koga}
The plasmon spectrum is very close to
that without SOI and it becomes difficult to distinguish one from the other.
On the other hand, compared to the single continuum of the 2DEG without SOI,
the single continuum with RSOI has an additional inter-branch band through which the plasmon
spectrum passes. The real part of the dielectric function at
$q=0.2\times 10^{-5}$ cm$^{-1}$ is plotted in Fig. \ref{fig3}(b),
whose zero near $\omega=7$ meV gives the plasmon energy
and zero near $\omega=1$ meV the edge of the intra-branch single continuum.
Because the spin of an electron in
the upper branch is exactly opposite to that
in the lower branch at the same wavevector ${\bf q}$,
the vertical inter-branch transition is not allowed under
the spin-independent Coulomb interaction,
there is no zero corresponding to the edge of the inter-branch single continuum.
Non-vertical inter-branch transitions, where the final electron wavevectors are
different from the initial ones, are possible
but very weak
and have negligible effect to the plasmon spectrum for
$k_F \gg k_\alpha$ or high electron densities
as indicated in Eq. (\ref{ratio}) and illustrated in Fig. \ref{fig3}.
Nevertheless, since the plasmon spectrum passes through the inter-branch
single continuum as shown in
Fig. \ref{fig3}(a),
the weak inter-branch transitions increase the Landau damping of plasmons
and lead to a nonzero imaginary part of the dielectric function in the corresponding energy range.
In Fig. \ref{fig3}(c) the nonzero value of Im$[\epsilon]$ in the energy range $(6,8.3)$ meV
is attributed
to the inter-branch transitions
and is much smaller than the value of Im$[\epsilon]$ inside the intra-branch single
continuum in the energy range $(0,1)$ meV.

The effect of RSOI on the 2D plasmon spectrum illustrated above coincides
with that presented
in Ref. \onlinecite{ullr},
where the spectrum is obtained by the density-functional formalism and
found to be independent of the RSOI.
However, the result disagrees with the one obtained in Ref. \onlinecite{maga}, where
the optical conductivity is used to evaluate the plasmon spectrum and
one extra mode is observed
at the edges of the inter-branch single continuum.
The disagreement
originates from the different approaches employed.
In this paper, the dynamically screened Coulomb interaction is
evaluated to find the plasmon modes
and the vertical transitions between the two SOI branches are not allowed under the bare
Coulomb interaction.
As a result, the plasmon spectrum
is not greatly affected by the SOI splitting. Similarly, the independence of
the plasmon spectrum of the RSOI found in Ref. \onlinecite{ullr} is a result of
the fact that the charge density matrix elements vanish for vertical transitions.
However, the nonvanishing matrix elements of the velocity operator
$\hat{\bf v}=\nabla_{\hat{\bf p}} \hat{H}$ corresponding to vertical transitions
introduces a new singularity
to the optical conductivity. Consequently, a new plasmon mode is predicted in
Ref. \onlinecite{maga} by an analytic treatment valid in the collision-free limit
and involving order-of-magnitude estimates,
see Eqs. (13) and (14) in Ref. \onlinecite{maga}.
Since the validity of this formalism in the presence of RSOI
is ambiguous by itself, a direct judgement on it is beyond the range of this paper.
Nevertheless, experimental observations of the plasmon spectrum of electrons in a quantum well
with typical parameters as used in Fig. \ref{fig3} is possible to clarify the controversy.
\begin{figure}[tpb]
\vspace*{-4cm}
\includegraphics*[width=100mm]{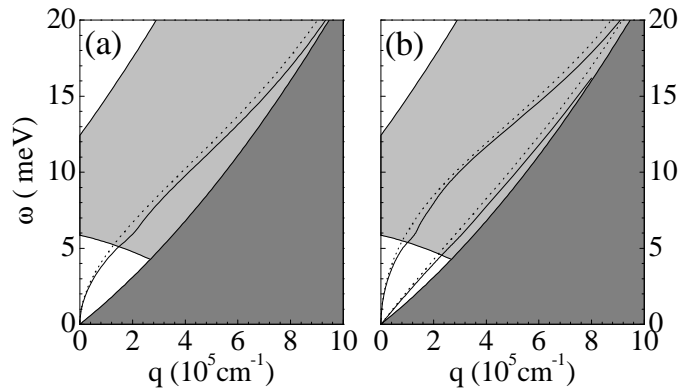}
\vspace{-3cm}
\caption{Plasmon spectrum of a 2DEG without SOI (dotted curve) and with RSOI
(solid curve) in a SQW (a) and in a DQW (b) .
The light (darker) shaded indicates the inter-branch (intra-branch) electron-hole continuum
in the presence of RSOI .
} 
\label{fig4}
\end{figure}

In Fig.\ref{fig4}(a), we plot the plasmon dispersion relation (solid curve)
 as well as 
the single particle excitation spectrum due to intra-branch (dark shadow area) and inter-branch (light grey area) transition for a 2DEG of electron density $n_e=1.5 n_0$
and RSOI strength $\alpha=5\alpha_0$. For the sake of comparison, the plasmon dispersion relation
in the absence of RSOI
is also shown (dotted curve).
We observe an overall lowering of the plasmon frequency in the presence of RSOI,
which results from the RSOI modification
of the band structure.
Different from the case without SOI where no overlap between the
plasmon and the single-excitation
spectra happens in the energy range shown in Fig.\ref{fig4}(a),
the plasmon spectrum with RSOI has a wider range of
overlap with the interbranch single-excitation
spectrum introduced by the RSOI spin split. Though numerical result shows that
the imaginary part of the dielectric function is thousand times smaller in the
inter-branch single-excitation regime than that in the range of
intra-branch regime, this energy overlap can enhance greatly the Landau damping of 2D plasmon with RSOI.
Near $q=2 \times 10^5$ cm$^{-1}$ where $\omega(q)\sim 2\alpha k^+_F$,
we notice also that a shoulder appears in the solid dispersion curve in Fig.\ref{fig4}(a).
Its explanation is given in the following.

\begin{figure}[tpb]
\vspace*{-3cm}
\includegraphics*[width=100mm]{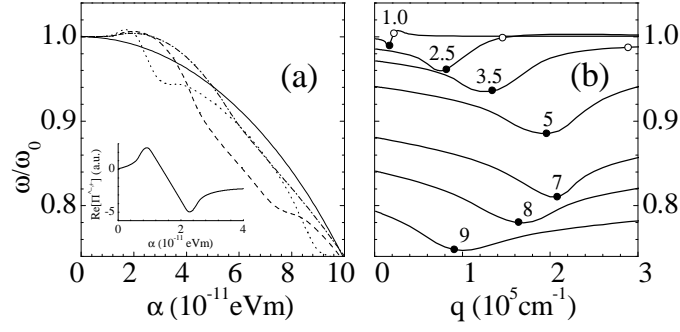}
\vspace{-4.5cm}
\caption{(a) Frequency ratio $\omega/\omega_0$ of a SQW plasmon
vs the RSOI strength
$\alpha$ for different wavevectors $q=0$ (solid curve), $10^5$ cm$^{-1}$ (dotted curve),
$2\times 10^5$ cm$^{-1}$ (dashed curve), and $3 \times 10^5$ cm$^{-1}$ (dot-dashed
curve). In the inset $\hat{\Pi}^{-,+}$ vs $\alpha$ for $q=0.1$ and $\omega=3$meV is shown. (b) $\omega/\omega_0$ vs $q$ with values of $\alpha$ in unit of $\alpha_0$ shown beside the curves.
The filled (empty) circles show the resonant positions $\omega= 2\alpha k^+_F$
($2\alpha k^-_F$). The electron density is $n_e=1.5n_0$ in both panels.}
\label{fig5}
\end{figure}

To have a closer look at the plasmon spectrum shift of RSOI, we define the
frequency ratio $\omega/\omega_0$ as the plasmon frequency in the presence of RSOI
divided by that in the absence of RSOI when other parameters are kept the same.
In Fig.\ref{fig5}(a), the frequency ratio $\omega/\omega_0$ is plotted as a function of
the RSOI strength $\alpha$ for various
wavevectors. In the long-wavelength limit $q=0$ (solid curve), only intra-branch transitions occour
and $\omega/\omega_0$
vs $\alpha$ follows Eq.(\ref{ratio}). For finite $q$ (dotted and dash-dotted curves),
inter-branch transitions also affect the plasmon spectrum especially
near
the inter-branch resonant energies at the Fermi wavevector of the $-$ branch, $\omega(q)\sim 2\alpha k^-_F$,
where an increase of plasmon frequency may appear, and of the $+$ branch, $\omega(q)\sim 2\alpha k^+_F$,
where a plasmon frequency decreases and a spectrum shoulder appears as shown in Fig. \ref{fig4}(a).
This behavior can be understood by checking the inter-branch correlation term in Eq. (\ref{diel}).
For positive
$\omega$, $\hat{\Pi}^{+,-}({\bf q},\omega)$ is negligible
and $\hat{\Pi}^{-,+}({\bf q},\omega)$ is
approximately expressed as
\begin{equation}
\hat{\Pi}^{-,+}({\bf q},\omega)=
\sum_{\bf k}
|g^{-,+}({\bf k},{\bf k}-{\bf q})|^2
\Pi^{-,+}_{\bf k}({\bf q},\omega)
\propto q^2k_\alpha/[\tilde{k}_F(\hbar\omega-2\alpha \tilde{k}_F)],
\label{inter}
\end{equation}
with $k_F^+<\tilde{k}_F<k_F^-$. A numerical result of $\Pi^{-,+}_{\bf k}({\bf q},\omega)$ as a function
of $\alpha$ is shown in the inset of Fig.\ref{fig5}(a),
which present smoothed peaks compared with that given by Eq.(\ref{inter}).
For strong RSOI, i.e. $2\alpha k_F^+ > \omega$,
$\Pi^{-,+}$ is positive and leads to an increase of plasmon frequency. If this positive
inter-branch correlation is larger than the RSOI negative extra term of the intra-branch correlation in Eq.(\ref{intra}),
there will be
a RSOI enhancement of the plasmon frequency. This enhancement is observed in Fig.\ref{fig5}(a)
near $\alpha=2\alpha_0$ with $\omega/\omega_0>1$.
For $2\alpha k_F^- < \omega$, $\Pi^{-,+}$ becomes negative and results in
a further lowering of plasmon frequency.
At fixed $q$, the frequency lowering of the inter-branch correlation appears generally
stronger than the frequency enhancement effect
because it happens at larger values of $\alpha$
and $\Pi^{-,+}$ is proportional to $\alpha$.
In Fig. \ref{fig5}(a), the curve of $q=10^5$ cm$^{-1}$ has two dips near $\alpha=3\alpha_0$ and $9\alpha_0$ respectively because there are resonant energies
$\omega\simeq E^+_{k^+_F}-E^-_{k^+_F}=2\alpha k_F^+$ here.

Since the plasmon frequency $\omega$ in Eq.(\ref{inter}) depends on the wavevector $q$,
the frequency ratio $\omega/\omega_0$ vs $q$ shows similar frequency enhancement and lowering as
illustrated in Fig.\ref{fig5}(b). The effect of RSOI via intra-branch correlation is reflected in
the value of $\omega/\omega_0$ at $q=0$ and the contribution via inter-branch correlation is observed in
the dependence on $q$. There is a lowering of the frequency ratio at low $q$ and a frequency enhancement
for high $q$. Each $\omega/\omega_0$ curve reaches its minimum (maximum)
at $q$ about $\omega(q)=2\alpha k_F^+$ ($2\alpha k_F^-$).
Among the curves in Fig. \ref{fig5}(b), the effect of this inter-branch correlation
is maximum for the curve $\alpha=7\alpha_0$
whose minimum happens at the maximum value of $q$.

\begin{figure}[tpb]
\vspace*{-3cm}
\includegraphics*[width=100mm]{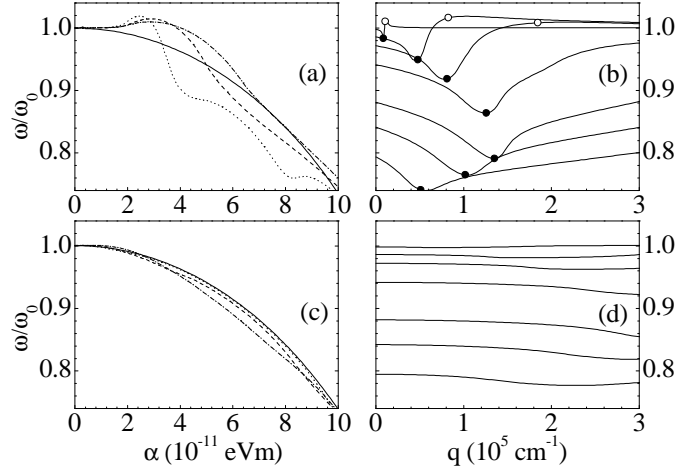}
\vspace*{-3.5cm}
\caption{Frequency ratios $\omega/\omega_0$ in a DQW as functions of
$\alpha$ and $q$
for the optic plasmon in (a) and (b), and the acoustic plasmon
in (c) and (d).
In panels (a) and (c), $\omega/\omega_0$ vs $\alpha$ for $q=0$ (solid),
$10^5$ cm$^{-1}$ (dotted), $2 \times 10^5$ cm$^{-1}$ (dashed),
and $3 \times 10^5$ cm$^{-1}$ (dot-dashed) are shown.
In panels (b) and (d) $\omega/\omega_0$ vs $q$ for (from up to down)
$\alpha/\alpha_0=1, 2.5, 3.5, 5, 7, 8, 9$ are shown.
The filled (empty) circles in (b) show the resonant positions as in Fig.\ref{fig5}(b).
The same electron density as in Fig.\ref{fig5} is used.}
\label{fig6}
\end{figure}

\subsection{Plasmons in a double well}

In the DQW case, we present the results of a structure composed of
two identical infinitely high quantum wells of
width $b=23$\AA\ separated by a distance $d=175$\AA.
The intra-well Coulomb interaction is $v_{11}(q)=v_{22}(q)=v(q)$ and the inter-well one 
$v_{12}(q)=v_{21}(q)=v_d(q)=e^2e^{-qd}G(qb)/(2\epsilon_0\epsilon_i q)$ with
$G(\xi)=16\pi^4(1-e^\xi)e^{-\xi}/[\xi^2(\xi^2+\pi^2)^2]$.
The plasmon spectrum is given by the zeros of the determinant of the dielectric matrix.
From Eq.(\ref{diel}), the determinant is expressed as
\begin{equation}
|\hat{\epsilon}|=(1-v\hat{\Pi}_1)(1-v\hat{\Pi}_2)-v_d^2\hat{\Pi}_1\hat{\Pi}_2
\end{equation}
with $\hat{\Pi}_i$ the noninteracting correlation function
in well $i$.
In general, two plasmon modes, the optic mode of higher energy and the acoustic mode,
are obtained out of the single
particle excitation spectrum.
If the electron densities in both wells are the same, Eq. (\ref{plsm}) is still
valid in the absence of RSOI for the zero-temperature plasmon dispersion
by replace $\Delta$ with
$\Delta_{op}=v+v_d$ for the optic mode and $\Delta_{ac}=v-v_d$ for the acoustic mode. 
In the long-wavelength limit, the frequency of the optic mode is about
$\omega_p^{op}=\sqrt{2}\omega_p$ and that of the acoustic mode 
$\omega_p^{ac}=(n_ee^2d/2\epsilon_0\epsilon_im^\ast)^{1/2}q$ .

In the numerical calculation,
the plasmon dispersion is obtained by obtaining the zeros of the real part of the determinant
of the dielectric matrix $Re|\hat{\epsilon}(q,\omega)|$. A typical plasmon dispersion of 2DEG
in a DQW with
RSOI is illustrated in Fig.\ref{fig4}(b).
As in the SQW, the dispersion curve of each plasmon mode is generally lowered by RSOI
via intra-branch correlations
and has a shoulder at the resonant energy
$2\alpha k_F^+$ of inter-branch transitions for electrons near the inner Fermi circle.
Compared with the shoulder of the SQW plasmon spectrum in Fig.\ref{fig4}(a),
the shoulder for the optic mode
happens at a lower wavevector by a factor
$1/\sqrt{2}$ and has an enhanced amplitude but that for the acoustic mode is almost dispensable.
We plot the RSOI strength dependence of the frequency ratio for the optic mode
in Fig.\ref{fig6}(a) and for the acoustic mode in Fig.\ref{fig6}(c). Both modes have the same dependence on $\alpha$
in the long wavelength limit $q=0$ as expressed by Eq.(\ref{ratio}). At finite wavevector, the frequency ratio shows strong deviation from that of $q=0$ for the optic mode but it is not the case for the acoustic mode. To see more clearly the dependence of $\omega/\omega_0$ on $q$, the DQW frequency ratio
$\omega/\omega_0$ vs 
wavevector $q$ for optic and acoustic modes are plotted in Fig.\ref{fig6}(b) and (d) respectively
with RSOI strength $\alpha=1,2.5,3.5,5,7,8,9\alpha_0$ for curves counted from top.
The curves of the optic mode are similar to those of the SQW plasmon with minima near
$\omega(q)=2\alpha k_F^+$ and maxima near $2\alpha k_F^-$. Because the DQW optic mode has a higher energy
than the SQW mode, the minima shift to lower $q$ values by a factor of $1/\sqrt{2}$.
The frequency ratio of the DQW acoustic plasmon mode, however, has a very weak dependence
on wavevector. Inter-branch transitions can lower the optic plasmon frequency
more than ten percent and the acoustic plasmon frequency less than three percent.

\begin{figure}[tpb]
\vspace*{-3cm}
\includegraphics*[width=100mm]{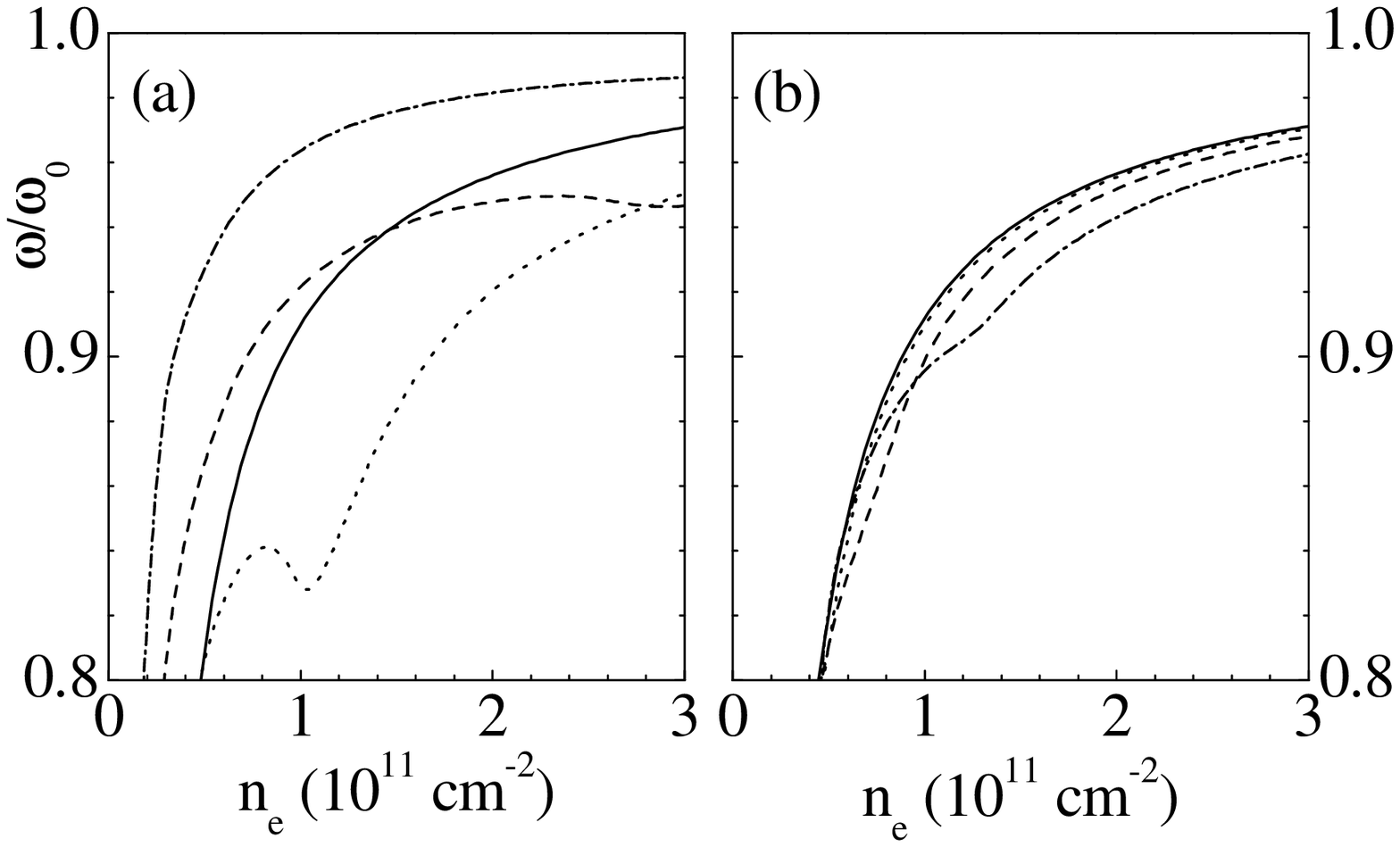
}
\vspace*{-3cm}
\caption{$\omega/\omega_0$ of (a) the optic and (b) the acoustic plasmons vs
$n_e$ in a DQW with $\alpha/\alpha_0=5$. Results of $q=0$ (solid),
$10^5$ cm$^{-1}$ (dotted), $2 \times 10^5$ cm$^{-1}$ (dashed), and $3 \times 10^5$ cm$^{-1}$ (dot-dashed)
are shown.} 
\label{fig7}
\end{figure}

The electron density dependence of the frequency ratio
$\omega/\omega_0$ is plotted for the optic mode in Fig. \ref{fig7}(a)
and the acoustic mode in Fig. \ref{fig7}(b) in a wavevector range up to $q=3\times 10^5$cm$^{-1}$.
In the long-wavelength limit $q=0$, it is described by Eq.(11) for both modes.
Fine structure appears in finite $q$ as inter-branch transitions are switched on.
For the optic mode, with increasing wavevector, the ratio decreases at first and a dip appear about
$\omega=2\alpha k_F^+$ which is $n_e=n_0$ for $q=10^{5}$cm$^{-1}$. Then the ratio increases from the
low-density regime to the higher-density regime. At $q=3$ it is higher than the value of $q=0$ over the whole regime.
For the acoustic mode, on the other hand, the ratio $\omega/\omega_0$ decreases and then increases
with $q$ in the low-density regime but keeps decreasing up to $q=3$ in the higher density regime.
The overall ratio at finite wavevectors
is lower than that in the zero-wavevector limit.
For $n_e > 3n_0$, all the curves $\omega/\omega_0$ in Fig. (\ref{fig7}) are above
$0.94$ and converge
to 1 with increasing $n_e$.

Note that our results are obtained in the RPA which may break down in systems
of low electron density or systems at high temperature.
The validity of the RPA can be justified by the parameter $r_s$ defined by
$n_e^{-1}=\pi r_s^2 a^{\ast 2}_B$ with the effective Bohr radius
$a^\ast_B=4\pi\epsilon_0\epsilon_\infty \hbar^2/(m^\ast e^2)$.
For a 2DES of electron density $n_0$, which corresponds to $r_s \approx 1.6$ here,
the RPA is still valid at low temperature.
Experiments also indicate that the RPA describes the plasmon spectrum and the Coulomb drag
of 2DEG's reasonably well at low temperature in GaAs/AlGaAs systems with electron density
of order $n_0$. \cite{kain,hill,wang}
For a system of lower electron density or at higher temperature, the local-field correction to the
dielectric function and the plasmon spectrum of 2DEG's becomes more important. \cite{hill,rojo,zhen,das1}

\section{Concluding remarks}
In the framework of the random phase approximation, we studied
the dielectric function and the plasmon spectrum
of single and coupled 2DEG's with RSOI in GaInAs-based semiconductor structures. In the SQW case,
the RSOI lifts the
spin degeneracy of 
electrons of finite wavevector
but does not change the spatial wave functions when the spin orientation of a electron varies.
Thus the bare Coulomb coupling between the electrons is not affected by the RSOI though
the spin effect to this
coupling should be considered explicitly in the presence of RSOI.
The screened Coulomb coupling and the dielectric function, nevertheless, are modified by RSOI
because the electron correlation in the system varies with the energy-band shape.
As a result of the invariance of spatial wavefunction of electrons in different spin branches,
the dielectric function
describing the Coulomb coupling between electrons in different spin branches has the same form
and the system is a single-component system.
The deformation of the energy band from a parabola for each spin branch
results in a negative term of intra-branch contributions to the dielectric function, which
lowers the plasmon frequency. 
The frequency decrease is proportional to the square of the RSOI strength and the inverse
of the electron density in the long wavelength limit. At finite wavevector of plasmon,
the contribution of inter-branch transitions to the dielectric function
is negative at energy lower than the spin split of the inner Fermi circle
and positive at energy higher than the spin split of
the outer Fermi circle. This leads to a shoulder
in plasmon dispersion curve at the energy equal to the spin split of inner Fermi circle and a
possible RSOI frequency enhancement for plasmon of higher energy dependent on the
competition between the intra- and inter-branch contributions.


In DQW systems, a similar influence of RSOI via intra-branch transitions is observed
on both the optic and acoustic modes. 
The effect of RSOI via inter-branch transitions is enhanced (diminished) for the optic
(acoustic) mode. The inter-branch single-particle excitation
spectrum covers part of the plasmon spectrum and leads to the Landau damping of plasmon.

\leftline{\bf Acknowledgments}
The author thanks P. Vasilopoulos and W. Xu for helpful discussions.
This work was supported by the Canadian NSERC Grant No. OGP0121756.


\begin{references} 
\bibitem{dat}  S. Datta and B. Das, Appl. Phys. Lett. {\bf 56}, 665 (1990).

\bibitem{wan} X. F. Wang, P. Vasilopoulos, and F. M. Peeters, Phys. Rev. B {\bf 65}, 165217 (2002).

\bibitem{dres}G. Dresselhaus, Phys. Rev. {\bf 100}, 580 (1955).

\bibitem{byc} Y. A. Bychkov and E. I. Rashba,
 JETP Lett. {\bf 39}, 78 (1984).

\bibitem{ver}L. Vervoort, R. Ferreira, and P. Voisin, Phys. Rev. B {\bf 56}, 12744 (1997).

\bibitem{ros} U. Rossler and J. Kainz, Solid State Commun. {\bf 121}, 313 (2002).

\bibitem{koga}see for example, J. Luo, H. Munekata, F. F. Fang, and P. J. Stiles, Phys. Rev. B {\bf 38}
10142 (1988);
T. Koga, J. Nitta, T. Akazaki, and H. Takayanagi, Phys. Rev. Lett. {\bf 89}, 046801 (2002).


\bibitem{chen} G. H. Chen and M. E. Raikh, Phys. Rev. B {\bf 59}, 5090 (1999).

\bibitem{maga}L. I. Magarill, A. V. Chaplik, and M. V. \'{E}ntin,
 JETP {\bf 92}, 153 (2001).
 
\bibitem{ullr}C. A. Ullrich and M. E. Flatt\'{e}, Phys. Rev. B {\bf 66}, 205305 (2002).

\bibitem{mish} E. G. Mishchenko and B. I. Halperin, Phys. Rev. B {\bf 68}, 45317 (2003).

\bibitem{xu} W. Xu, Appl. Phys. Lett. {\bf 82}, 724 (2003).


\bibitem{das} S. Das Sarma and A. Madhukar, Phys. Rev. B {\bf 23}, 805
(1981).

\bibitem{san} G. E. Santoro and G. F. Giuliani, Phys. Rev. B {\bf 37}, 937
(1988).

\bibitem{fas}G. Fasol, N. Mestres, H. P. Hughes, A. Fischer, and K. Ploog,
Phys. Rev. Lett. {\bf 56}, 2517 (1986).

\bibitem{kain} D. S. Kainth, D. Richards, A. S. Bhatti, H. P. Hughes, M.
Y. Simmons, E. H. Linfield, and D. A. Ritchie, Phys. Rev. B {\bf 59},
2095 (1999).

\bibitem{hu} C.-M. Hu, C. Sch\"{u}ller, and D. Heitmann, Phys. Rev. B {\bf 64}, 073303 (2001).

\bibitem{flen}K. Flensberg and BenYu-Kuang Hu, Phys. Rev. Lett. {\bf 73},
3572 (1994).

\bibitem{hill}N. P. R. Hill, J. T. Nicholls, E. H. Linfield, M. Pepper,
D. A. Ritchie, G. A. C. Jones, BenYu-Kuang Hu, and K. Flensberg, Phys. Rev.
Lett. {\bf 78}, 2204 (1997).

\bibitem{rojo}A. G. Rojo, J. Phys.: CM {\bf 11}, R31 (1999).

\bibitem{wang}X. F. Wang and I. C. da Cunha Lima, Phys. Rev. B {\bf 63},
205312 (2001).

\bibitem{vin} B. Vinter, Phys. Rev. B {\bf 15}, 3947 (1977).

\bibitem{zhen} L. Zheng and A. H. MacDonald, Phys. Rev. B {\bf 49}, 5522 (1994).
\bibitem{das1} S. Das Sarma and E. H. Hwang, Phys. Rev. Lett. {\bf 81}, 4216 (1998).
\end{references}
\end{document}